\documentclass[aps,prd,showpacs,showkeys,floatfix,nofootinbib,superscriptaddress]{revtex4}
\usepackage{dcolumn}

\usepackage{amsmath,amsfonts,amssymb,mathrsfs,times,bm}
\usepackage{extarrows}
\usepackage{graphicx}
\usepackage{float}
\usepackage{graphicx,epstopdf}
\usepackage{dcolumn}
\usepackage{exscale}
\usepackage{relsize}
\usepackage{mathtools}
\usepackage[colorlinks=true,breaklinks=true,linkcolor=blue,citecolor=blue,urlcolor=blue]{hyperref}

\usepackage{allpurpose}

\usepackage{makeidx}
\usepackage[usenames,dvipsnames]{pstricks}
\usepackage{subfigure}
\usepackage{epsfig}
\usepackage{pst-grad} 
\usepackage{array}
\usepackage{tikz}
\usetikzlibrary{decorations.markings,math,shapes.misc,patterns,arrows.meta,calc,patterns,angles,quotes}

\newcommand{\nn}{\nonumber}

\begin{document}

\title[]
{The finite-distance gravitational deflection of massive particles in stationary spacetime: a Jacobi metric approach}
\author{Zonghai Li}
\affiliation{MOE Key Laboratory of Artificial Micro- and Nano-structures, School of Physics and Technology, Wuhan University, Wuhan, 430072, China}

\author{Junji Jia}
\email[Corresponding author email:~]{junjijia@whu.edu.cn}
\affiliation{Center for Astrophysics \& MOE Key Laboratory of Artificial Micro- and Nano-structures, School of Physics and Technology, Wuhan University, Wuhan, 430072, China}
\date{\today}

\begin{abstract}
In this paper, we study the weak gravitational deflection of relativistic massive particles for a receiver and source at finite distance from the lens in stationary, axisymmetric and asymptotically flat spacetimes. For this purpose, we extend the generalized optical metric method to the generalized Jacobi metric method by using the Jacobi-Maupertuis Randers-Finsler metric. More specifically, we apply the Gauss-Bonnet theorem to the generalized Jacobi metric space and then obtain an expression for calculating the deflection angle, which is related to Gaussian curvature of generalized optical metric and geodesic curvature of particles orbit. In particular, the finite-distance correction to the deflection angle of signal with general velocity in the the Kerr black hole and Teo wormhole spacetimes are considered. Our results cover the previous work of the deflection angle of light, as well as the deflection angle of massive particles in the limit for the receive and source at infinite distance from the lens object. In Kerr black hole spacetime, we compared the effects due to the black hole spin, the finite-distance of source or receiver, and the relativistic velocity in microlensings and lensing by galaxies. It is found in these cases, the effect of BH spin is usually a few orders larger than that of the finite-distance and relativistic velocity, while the relative size of the latter two could vary according to the particle velocity, source or observer distance and other lensing parameters. 


\end{abstract}

\maketitle

\section{Introduction}
100 years ago, Eddington~\textit{et. al}~\cite{DED1920,Will2015} firstly verified the general relativity through the deflection of light in the solar gravitational field. Nowadays gravitational lensing (GL) becomes a powerful tool in astrophysics and cosmology. For examples, it is used to measure the mass of galaxies and clusters~\cite{Hoekstra2013,Brouwer2018,Bellagamba2019} and to to detect dark matter and dark energy~\cite{Vanderveld2012,cao2012,zhanghe2017,Huterer2018,SC2019,Andrade2019}.

One of the main quantities in the study of GL is the deflection angle. Various approaches relying on the geodesics were built to calculate it. In 2008, Gibbons and Werner~\cite{GW2008} proposed a geometrical and topological method of studying the gravitational deflection of light in a static and spherically symmetric spacetime using the Gauss-Bonnet (GB) theorem. Later, Werner~\cite{Werner2012} extended this method to the rotating and stationary spacetimes by using the Randers-Finsler geometry. In Gibbons-Werner method, the deflection angle can be obtained by integrating the Gaussian curvature of corresponding optical metric in an infinite region enclosed by the geodesics and an infinitely large boundary. This method is fundamentally different from the standard geodesics approach. The importance of Gibbons-Werner method is that it shows the deflection angle can be viewed as a global topological effect. In addition, by this approach we only work with spatial geodesic in the two-dimensional positive definite Riemannian space, instead of the null or timelike geodesic in background spacetime, and thus the physical lens models can be implemented easily~\cite{Werner2012}. By using the Gibbons-Werner method, many authors studied the gravitational deflection angle of light not only for black holes (BHs), but for other lens object such as wormholes, and for both  asymptotically flat spacetimes and non-flat spacetimes such as a spacetime with cosmic string ~\cite{Jus-175,Jus-183,OV2018,Goulart2018,Javed2019,AO2019,LZ2019} .

On the other hand, massive particles such as neutrinos \cite{Hirata:1987hu, Bionta:1987qt} and potentially gravitational waves (GW)  \cite{TheLIGOScientific:2017qsa,GBM:2017lvd,Monitor:2017mdv}  in some modified gravitational theories can also be messengers in GLs. Correspondingly, people also studied the gravitational deflections of massive particles due to its importance such as studying the properties of massive neutrinos, gravitational wave and cosmic rays. In fact, the gravitational deflection of the massive particles has been studied in different spacetime with great interest ~\cite{AR2002,AP2004,Bhadra2007,Yu2014,He2016,He2017a,He2017b,Jia2016,Jia2019}.
A question naturally came: Can one apply the GB theorem to study the deflection of massive particles? The answer was first given by Crisnejo and Gallo~\cite{CG2018} which studied not only the deflection of light moving in plasma medium but also the deflection of massive particles in static and spherically symmetric spacetimes. In addition, Ref.~\cite{LHZ2019} studied the deflection of massive particles in static wormhole spacetimes according to Jacobi metric and GB theorem. Recently, Jusufi~\cite{Jus-massive1} studied the deflection of massive particles in stationary and axisymmetric spacetime using GB theorem. Moreover, the deflection angles of massive particles were used to distinguish rotating naked singularities from Kerr-like
wormholes in Ref.~\cite{Jusufi2019-1}. In addition, Refs.~\cite{CGV2019,Jusufi2019-2} studied the deflection of massive charged particles by charged BH using the GB theorem. Very recently, Crisnejo \textit{et al} extended their study in Ref.~\cite{CG2018} to the stationary spacetimes~\cite{CGJ2019}.

Typically, in most calculations involving the GB theorem, the weak-field limit is considered for the receiver and source at infinite distance from a lens object. However, in reality they are always located at a finite distance. The work of Gibbons and Werner allows some authors to take account the finite distance of the receiver and source into the gravitational deflection of light. By using the GB theorem, Ishihara \textit{et al} studied the finite-distance deflection of light in static and spherically symmetric spacetime both in weak~\cite{ISOA2016} and strong~\cite{IOA2017} field limits. Along the same line, Ono \textit{et al}~\cite{OIA2017,OIA2018} proposed the generalized optical metric method and used it to study the finite-distance deflection angle of light in stationary, axisymmetric and asymptotically flat spacetime. Their method was extended to stationary and non-asymptotically flat spacetime such as a rotating global monopole quite recently~\cite{OIA2019}. Furthermore, these authors considered the finite-distance correlation of the deflection angle of light and described its possible astronomical application due to the deflection of light in the solar gravitational field and Sgr $A^\ast$ gravitational field. In addition, a review on finite-distance deflection of light was given by Ono and Asada in Ref.~\cite{OA2019}. It
is worthwhile to mention that another work for finite-distance deflection of light in static and spherically symmetric spacetime was established by Arakida~\cite{Arakida2018}. However, the results are different from Ref.~\cite{ISOA2016} and Ref.~\cite{Arakida2018}. Recently, Crisnejo and Gallo~\cite{CG2019} clarified this difference and further studied the finite-distance deflection of light in spherically symmetric gravitational field with a plasma medium.

This paper will study the finite-distance gravitational deflection of massive particles in the stationary, axisymmetric and asymptotically flat spacetime. In particular, we consider the Kerr BH spacetime and Teo wormhole spacetime in details. In the previous works, the infinite-distance deflection of light~\cite{Werner2012,Jus-183} and massive particles~\cite{Jus-massive1,CGJ2019}, and finite-distance deflection of light~\cite{OIA2017,OIA2018} in these two spacetime were studied in the weak field limits using the GB theorem. In this paper, we shall extend the study to the finite-distance gravitational deflection of the relativistic massive particles in the weak field limits and compare the effects of spacetime spin, finite distance and subluminal velocity in the microlensing and supermassive BH lensing cases. To this end, we mainly consider the generalized optical metric method~\cite{OIA2017,OIA2018}. In order to use the GB theorem,  the study of light deflection is carried out in the optical metric space~\cite{GW2008}, whereas study of particle deflection is done in the Jacobi (Jacobi-Maupertuis) metric space~\cite{Gibbons2016}. Both the corresponding optical and Jacobi metrics of stationary spacetimes are indeed Randers-Finsler metrics. Therefore, this paper will use Jacobi-Maupertuis Randers-Finsler (JMRF) metric~\cite{Chanda2019} rather than optical Randers-Finsler (ORF) metric and we need to extend the generalized optical metric method to the generalized Jacobi metric method first.

This paper is organized as follows. In Section~\ref{Jacobi} , we review the JMRF metric and derive the orbit equation of massive particles in stationary and axisymmetric spacetimes, and then extend the generalized optical metric method to the generalized Jacobi metric method using the JMRE metric. In section~\ref{Kerr}, we study the deflection angle of massive particles in Kerr BH spacetime for a receiver and source at finite distance using the generalized Jacobi metric method. With the same process, Section~\ref{TEO} computes the finite-distance deflection angle of massive particles in rotating Teo wormhole spacetime. Finally, we comment on our results in Section~\ref{CONCLU}. Throughout this paper, we take the unit of $G = c = 1$ and the spacetime signature $(+,-,-,-)$.

\section{JMRF metric and the generalized Jacobi metric method}\label{Jacobi}
\subsection{JMRF metric }
 In this subsection we review the JMRF metric derived by Chanda \textit{et al} in Ref.~\cite{Chanda2019}. We begin by the line element for an arbitrary stationary spacetime
\begin{eqnarray}
&& ds^2=g_{tt}(\pmb{x})dt^2+2g_{ti}(\pmb{x})dt dx^i+g_{ij}(\pmb{x})dx^i dx^j~, \label{eq:mastermetric}
\end{eqnarray}
where $\pmb{x}$ is the spatial coordinates.
Then, the corresponding relativistic Lagrangian for a free particle reads
\begin{eqnarray}
 &&\mathcal{L}=-m\sqrt{g_{\mu \nu} \dot{x}^{\mu} \dot{x}^{\nu}}=-m\sqrt{g_{tt}\dot{t}^2+2g_{ti}\dot{t}\dot{x}^i+g_{ij}\dot{x}^i \dot{x}^j}~,
\end{eqnarray}
where $m$ is the particle mass and the dot denotes the differentiation with respect to an arbitrary parameter. Then, one can write the canonical momentum as
\begin{eqnarray}
\label{canonical momentum1}
&& p_t=\frac{\partial \mathcal{L}}{\partial \dot{t}}=-m\frac{g_{tt}\dot{t}+g_{ti}\dot{x}^i}{\sqrt{g_{\alpha\beta}\dot{x}^\alpha\dot{x}^\beta}}=-\mathcal{E}~,\label{eq:ptdef}\\
\label{canonical momentum2}
&& p_i=\frac{\partial \mathcal{L}}{\partial \dot{x}^i}=-m\frac{g_{ti}\dot{t}+g_{ij}\dot{x}^j}{\sqrt{g_{\alpha\beta}\dot{x}^\alpha\dot{x}^\beta}}~,
\end{eqnarray}
where $\mathcal{E}=p_i-\mathcal{L}$ is the relativistic energy of the particle.
Now, the Jacobi Lagrangian reads
\begin{eqnarray}
\label{Jacobi Lagrangian}
 L_J&=&p_i \dot{x}^i=-m\frac{g_{ti}\dot{t}\dot{x}^i+g_{ij}\dot{x}^i \dot{j}^j}{\sqrt{g_{\alpha\beta}\dot{x}^\alpha\dot{x}^\beta}} =p_t\frac{g_{ti}}{g_{tt}}\dot{x}^i+m\sqrt{\frac{\gamma_{ij}\dot{x}^i \dot{x}^j}{g_{\alpha\beta}\dot{x}^\alpha \dot{x}^\beta}}\sqrt{\gamma_{ij}\dot{x}^i \dot{x}^j}~,~~~
\end{eqnarray}
where Eq. \eqref{eq:ptdef} was used and the spatial metric $\gamma_{ij}$ is defined as
\begin{eqnarray}
\label{spatial metric}
&& \gamma_{ij}:=-g_{ij}+\frac{g_{ti} g_{tj}}{g_{tt}}~.~~~
\end{eqnarray}
Taking square of Eq. \eqref{eq:ptdef} and using Eqs \eqref{spatial metric} and \eqref{eq:mastermetric}, we have
\begin{eqnarray}
&& p_t^2=m^2g_{tt}\left(1+\frac{\gamma_{ij}\dot{x}^i \dot{x}^j}{g_{\alpha\beta}\dot{x}^\alpha \dot{x}^\beta}\right),
\end{eqnarray}
from which we can solve
\begin{eqnarray}
&&\frac{\gamma_{ij}\dot{x}^i \dot{x}^j}{g_{\alpha\beta}\dot{x}^\alpha \dot{x}^\beta}=\frac{p_t^2-m^2 g_{tt}}{m^2 g_{tt}}~.
\end{eqnarray}
Substituting this into Eq.~\eqref{Jacobi Lagrangian}, one has
\begin{eqnarray} L_J&=&F\left(\pmb{x},\dot{\pmb{x}}\right)=p_t\frac{g_{ti}}{g_{tt}}\dot{x}^i+\sqrt{\frac{p_t^2-m^2 g_{tt}}{ g_{tt}} \gamma_{ij}\dot{x}^i \dot{x}^j}~,~~~
\end{eqnarray}
and the Jacobi-Maupertuis metric can be written as~\cite{Chanda2019}
\begin{eqnarray}
\label{Randers-Finsler}
ds_J&=&p_idx^i=\sqrt{\frac{\mathcal{E}^2-m^2 g_{tt}}{g_{tt}}\gamma_{ij}dx^i dx^j}-\mathcal{E} \frac{g_{ti}}{g_{tt}}dx^i
\equiv \sqrt{\alpha_{ij}dx^i dx^j}+\beta_i dx^i~,
\end{eqnarray}
where we have used $p_t=-\mathcal{E}$. Eq.~\eqref{Randers-Finsler} implies that the Jacobi-Maupertuis metric $ds_J$ is a Finsler-Randers metric, which satisfies the positivity and convexity~\cite{Chern2002}
\begin{eqnarray}
&&\sqrt{\alpha^{ij}\beta_i \beta_j}<1~,
\end{eqnarray}
with $\alpha_{ij}$ being a Riemannian metric and $\beta_i$ being a one-form. Importantly, the trajectories of neutral particle moving in a stationary metric are seen as the geodesics of the corresponding Finsler-Randers metric space.

For $m=0$ and $\mathcal{E}=1$, Eq.~\eqref{Randers-Finsler} reduces to the ORF metric~\cite{Werner2012}
\begin{eqnarray}
ds_O&=&\sqrt{\left(-\frac{g_{ij}}{g_{tt}}+\frac{g_{ti} g_{tj}}{g_{tt}^2}\right)dx^i dx^j}-\frac{g_{ti}}{g_{tt}}dx^i~.
\end{eqnarray}
 For $g_{ti}=0$, Eq.~\eqref{Randers-Finsler} reduces to the Jacobi metric for static spacetime~\cite{Gibbons2016}
\begin{eqnarray}
&& ds_J^2=\frac{m^2 g_{tt}-\mathcal{E}^2}{g_{tt}}g_{ij}dx^i dx^j~.
\end{eqnarray}
\subsection{Motion of massive particle on the equatorial plane}
This paper will focus on the stationary, axisymmetric and asymptotically flat spacetimes (which certainly include the static and spherically symmetric spacetimes), and their line element in the polar coordinates $(t,r,\theta,\varphi)$ can be written as~\cite{OIA2017}
\begin{eqnarray}
\label{line element}
 ds^2&=&g_{tt}\left(r,\theta\right)dt^2+2g_{t\varphi}\left(r,\theta\right)dt d\varphi+g_{rr}\left(r,\theta\right)dr^2+g_{\theta\theta}\left(r,\theta\right)d\theta^2+g_{\varphi\varphi}\left(r,\theta\right)d\varphi^2~.
\end{eqnarray}
For simply, we only study the motion of particle on the equatorial plane $(\theta=\pi/2)$.
 Choose the appropriate parameter such that
 \begin{eqnarray}
\label{r1}
 1&=&g_{\mu \nu} \dot{x}^{\mu} \dot{x}^{\nu}=g_{tt}\dot{t}^2+2g_{t\varphi}\dot{t } \dot{\varphi}+g_{rr}\dot{r}^2+g_{\varphi\varphi}\dot{\varphi}^2~.
\end{eqnarray}
Then one can obtain two conserved quantities from Eqs.~\eqref{canonical momentum1} and~\eqref{canonical momentum2}
\begin{eqnarray}
\label{ELL}
&& m\left(g_{tt}\dot{t}+g_{t\varphi}\dot{\varphi}\right)=\mathcal{E}~,~-m\left(g_{t\varphi}\dot{t}+g_{\varphi \varphi}\dot{\varphi}\right)=\mathcal{J}~,
\end{eqnarray}
where $\mathcal{J}$ is the conserved angular momentum of the particle. They can be measured at infinity for an asymptotic observer by
\begin{eqnarray}
\label{EJ}
&& \mathcal{E}=\frac{m}{\sqrt{1-v^2}}~,~\mathcal{L}=\frac{mvb}{\sqrt{1-v^2}}~,
\end{eqnarray}
where $v$ is the particle velocity at infinity and $b$ is the impact parameter defined by
\begin{eqnarray}
\label{impact parameter}
&& bv\equiv\frac{\mathcal{J}}{\mathcal{E}}~.
\end{eqnarray}
Introducing the inverse radial coordinate $u=1/r$, the orbit equation of massive particles can be obtained from Eqs.~\eqref{r1}-~\eqref{EJ} as the following
\begin{eqnarray}
\label{Sorbit}
\left(\frac{du}{d\varphi}\right)^2&=&\frac{u^4\left(g_{t\varphi}^2-g_{tt}g_{\varphi\varphi}\right)\left[g_{tt}b^2v^2+2g_{t\varphi}b v+g_{t\varphi}^2\left(1-v^2\right)+g_{\varphi\varphi}\left(1-g_{tt}+g_{tt}v^2\right)\right]}{g_{rr}\left(g_{t\varphi}+g_{tt} b v\right)^2}~.~~~~
\end{eqnarray}
\subsection{The GB theorem}
Suppose that $D$ is a subset of a compact, oriented surface, with Gaussian curvature $K$ and Euler characteristic $\chi(D)$. Its boundary $\partial{D}$ is a piecewise smooth curve with geodesic curvature $k_g$. In the i-th vertex of $\partial{D}$, the jump angle is denoted as $\phi_i$ in the positive sense. Then, the GB theorem states that~\cite{GW2008,Carmo1976}:
\begin{equation}
\iint_D{K}dS+\oint_{\partial{D}}k_g~d\sigma+\sum_{i=1}{\phi_i}=2\pi\chi(D)~,\\
\end{equation}
where $dS$ is the area element of $D$ and $d\sigma$ is the line element along $\partial{D}$.

For infinite-distance case, one could apply the Werner's Finsler geometry method and the deflection angle can be computed by~\cite{Werner2012}
\begin{eqnarray}
\label{GBT-K}
\hat{\alpha}=-\iint_{D_{\infty}}{K}dS~,
\end{eqnarray}
where $D_{\infty}$ denotes  the infinite Jacobi region out of the particle trajectory.

However, the Finsler geometry is difficult to use for the calculation of the deflection angle for the receiver and source at finite distance from the lens. The reason is that the definition of jump angles at the vertices in the GB theorem are problematic in the Finsler geometry, and thus Ono \textit{et al} proposed the generalized optical metric method to avoid the Finsler geometry ~\cite{OIA2017, OIA2018}. The JMRF metric is quite parallel to the ORF metric, which allows us to apply the formulas in~\cite{OIA2017, OIA2018}.
\subsection{The generalized Jacobi metric method}
We call the positive Riemannian metric $\alpha_{ij}$ as generalized Jacobi metric and suppose that the particles live in the Remannian space $\bar{M}$ described by the generalized Jacobi metric
\begin{eqnarray}
\label{GJM}
&& dl^2=\alpha_{ij}dx^idx^j~.
\end{eqnarray}
Thus, the motion equation of particles can be written as~\cite{OIA2017}
\begin{eqnarray}
\label{me}
&& \frac{de^i}{dl}+\bar{\Gamma}_{jk}^{i}e^ie^j=\alpha^{ij}\left(\beta_{k\mid j}-\beta_{j\mid k}\right)e^k,
\end{eqnarray}
where $e^i\equiv dx^i / dl$ is the unit tangential vector along particle ray, the barred quantities means that they are related to the generalized Jacobi metric, and $\mid$ denotes the covariant derivative using $\alpha_{ij}$.

It is obvious from Eq.~\eqref{me} that the particle ray now is not the geodesic in $\bar{M}$, due to the existence of $\beta_i$. Hence, the geodesic curvature of particle orbit will not vanish in $\bar{M}$ and it can be calculated by~\cite{OIA2017}
\begin{eqnarray}
\label{gc1}
&& k_g=-\epsilon^{ijk}N_i\beta_{j\mid k}~,
\end{eqnarray}
where $\epsilon^{ijk}$ is the Levi-Civita tensor defined by $\epsilon^{ijk}\equiv \sqrt{\det \alpha }\varepsilon_{ijk}$ with $\varepsilon_{ijk}$ being Levi-Civita symbol, $N_i$ is the unit normal vector. For metric~\eqref{line element},  Eq.~\eqref{gc1} becomes~\cite{OIA2017}
\begin{eqnarray}
\label{gc2}
&& k_g=-\frac{\beta_{\varphi,r}}{\sqrt{{\det \alpha ~\alpha^{\theta\theta}}}}~,
\end{eqnarray}
where the comma denotes the partial derivative.

Now, we use the GB theorem to study the finite-distance deflection angle in the equatorial plane ($\theta=\pi/2$). 
First, we apply the definition of deflection angle~\cite{OIA2017}
\begin{eqnarray}
\label{angle}
&& \hat{\alpha}\equiv \Psi_R-\Psi_S+\varphi_{RS}~,
\end{eqnarray}
where $\Psi_R$ and $\Psi_S$ are angles between the particle ray's tangent and the radial direction from the lens to the receiver and source, respectively. The coordinate angle $\varphi_{RS}\equiv\varphi_R-\varphi_S$, where $\varphi_R$ and $\varphi_S$ are the angular coordinates of the receiver and source. 

\begin{figure}[htb!]
\centering
\includegraphics[width=8.5cm]{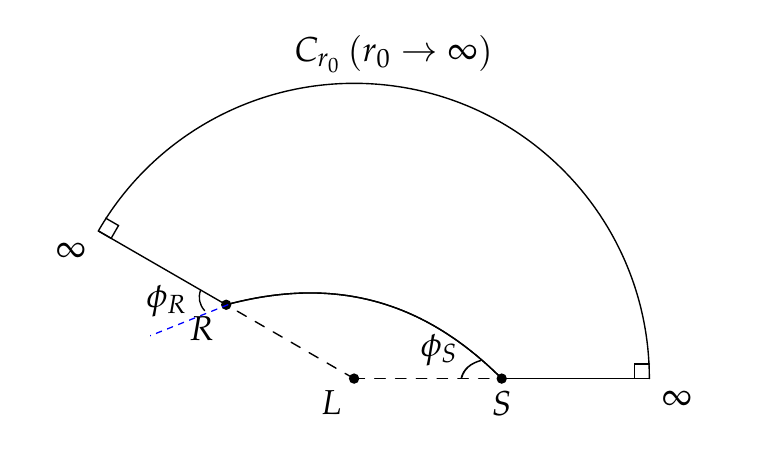}
\caption{The quadrilateral $\prescript{\infty}{R}\Box_{S}^{\infty}\subset(\bar{M},\alpha_{ij})$. R, L and S denote the receiver, the lens and the source, respectively. $\Psi_R$ and $\Psi_S$ are angles between the particle ray tangent and the radial direction from the lens in R and S, respectively. The curve $C_{r_0}$ is defined by $r(\varphi)=r_0=constant$. Note that each outer angle at the intersection of the radial direction curves and $C_{r_0}$ is $\pi/2$, as $r_0\rightarrow\infty$. Note that the jump angle $\phi_S=\pi-\Psi_S$ and $\phi_R=\Psi_R$.}\label{Figure}
\end{figure}

Following~\cite{OIA2017}, we consider the quadrilateral $\prescript{\infty}{R}\Box_{S}^{\infty}\subset(\bar{M},\alpha_{ij})$ as shown in Fig.~\ref{Figure}. It is bounded by four curves: the particle trajectory from source (S) to receiver (R), two spatial geodesics of outgoing radial lines passing through R and S respectively, and a circular arc segment $C_{r_0}(r_0\rightarrow \infty)$, where $C_{r_0}$ is defined by $r(\varphi)=r_0=$constant. 
For curve $C_{r_0}$, we have $k_gdl\rightarrow d\varphi$ when $r_0\rightarrow\infty$, because $\bar M$ is asymptotically flat space. Thus, we have $\lim\limits_{r_0\rightarrow\infty} \int_{C_{r_0}}k_gdl=\varphi_{RS}$. In addition, by the construction of $\prescript{\infty}{R}\Box_{S}^{\infty}$ one can see that its Euler characteristic is unit. Notice that the sum of two jump angles at infinite is $\pi$. In addition, we have $\phi_S=\pi-\Psi_S$ and $\phi_R=\Psi_R$. Finally, we use GB theorem to the quadrilateral and obtain
 {\small\begin{equation}
\iint_{_{R}^{\infty}\Box_{S}^{\infty}}K dS-\int_{S}^{R}k_gdl+\varphi_{RS}+\Psi_R-\Psi_S=0~.\\
\end{equation}}
 By this expression, Eq.~\eqref{angle} can be rewritten as
\begin{eqnarray}
\label{gbdef}
\hat{\alpha}=-\iint_{_{R}^{\infty}\Box_{S}^{\infty}}K dS+\int_{S}^{R}k_g dl~.
\end{eqnarray}
This expression clearly shows that the deflection angle $\hat{\alpha}$ is coordinate-invariant. One can calculate the Gaussian curvature of generalized Jacobi metric induced in the equatorial plane by~\cite{Werner2012}
{\small\begin{eqnarray}
\label{Gauss-K}
{K}&=&\frac{\bar{R}_{r \varphi r \varphi}}{\det{\alpha}}=\frac{1}{\sqrt{\det \alpha}}\left[\frac{\partial}{\partial{\varphi}}\left(\frac{\sqrt{\det \alpha}}{\alpha_{rr}}{{\bar{\Gamma}}^\varphi_{rr}}\right)-\frac{\partial}{\partial{r}}\left(\frac{\sqrt{\det \alpha}}{\alpha_{rr}}{{\bar{\Gamma}}^\varphi_{r\varphi}}\right)\right]~.~~~~~
\end{eqnarray}}

\section{Kerr BH deflection angle}\label{Kerr}
\subsection{Kerr-JMRF metric}
The Kerr BH is a stationary, axisymmetric and asymptotically flat solution for Einstein field equation. The line element of Kerr BH with mass $M$ and angular momentum per unit mass $a$ reads~\cite{BL1967}
{\small\begin{eqnarray}
\label{Kerrlm}
ds^2&=&\left(1-\frac{2M r}{\Sigma}\right)dt^2+\frac{4M a r \sin^2 \theta}{\Sigma}d\varphi dt-\frac{\Sigma}{\Delta}dr^2-\Sigma d\theta^2-\left(r^2+a^2+\frac{2 M a^2 r\sin^2 \theta}{\Sigma}\right)\sin^2\theta d\varphi^2~,~~~
\end{eqnarray}}
where
\begin{eqnarray}
&&\Sigma=r^2+a^2\cos^2\theta~.
\end{eqnarray}
Now, the spatial metric defined by Eq.\eqref{spatial metric} can be written as
\begin{eqnarray}
&& \gamma_{ij}^{\mathrm{K}}dx^i dx^j=\Sigma\left(\frac{dr^2}{\Delta}+d\theta^2+\frac{\Delta \sin^2 \theta}{\Delta-a^2\sin^2\theta}d\varphi^2\right)~.~~~~~~~
\end{eqnarray}
According to~\eqref{Randers-Finsler}, we find the following Kerr-JMRF metric
\begin{eqnarray}
\label{KerrJMRE}
\alpha_{ij}^{\mathrm{K}}dx^i dx^j&=&\left(\frac{{\mathcal{E}}^2\Sigma}{\Delta-a^2\sin^2\theta}-m^2\right)\Sigma\left(\frac{dr^2}{\Delta}+d\theta^2+\frac{\Delta \sin^2\theta}{\Delta-a^2\sin^2\theta}d\varphi^2\right)~,\nn\\
\beta_i^{\mathrm{K}} dx^i&=&-\frac{2\mathcal{E} M a r \sin^2\theta}{\Delta-a^2\sin^2\theta}d\varphi~,
\end{eqnarray}
which is firstly given in Ref.~\cite{Chanda2019}. For $m=0$ and $\mathcal{E}=1$, the Kerr-JMRF metric reduces to the Kerr-ORF metric~\cite{Werner2012}.
\subsection{Gaussian curvature}
Considering Eq.~\eqref{KerrJMRE}, we can find the generalized Kerr Jacobi metric in the equatorial plane $(\theta=\pi/2)$ as following
\begin{eqnarray}
\label{Kerrline}
dl^2&=&\alpha^{\mathrm{K}}_{ij}dx^idx^j=\left[\frac{r^2}{\left(r^2-2M r\right)\left(1-v^2\right)}-1\right]m^2r^2\left(\frac{dr^2}{\Delta}+\frac{\Delta d\varphi^2}{r^2-2M r}\right)~,~~~
\end{eqnarray}
where we have used Eq.~\eqref{EJ}. We will compute the deflection angle up to the second order. By Eq.~\eqref{Gauss-K}, we can obtain the corresponding Gaussian curvature up to second order given by
\begin{eqnarray} K_{\mathrm{K}}&=&-\frac{\left(1-v^4\right)M}{m^2r^3v^4}+\frac{3\left(2-3v^2+v^4\right)M^2}{m^2r^4v^6}+\mathcal{O}(M^3,M^2a,a^2M,a^3)~,
\end{eqnarray}
where the terms containing $a$ are more than second order in $K_{\mathrm{K}}$.
Considering Eq.~\eqref{Kerrlm} and Eq.~\eqref{Sorbit}, one can obtain the solution of orbit equation by perturbation method as
\begin{eqnarray}
\label{0th}
u(\varphi)&=&\frac{\sin \varphi}{b}+\frac{1+v^2\cos^2 \varphi}{b^2v^2}M-\frac{2aM}{b^3v}+\mathcal{O}(M^2,a^2)~.~~~~
\end{eqnarray}
In addition, we can obtain the iterative solution for $\varphi$ in the above equation as
\begin{eqnarray}
\varphi=\begin{cases}
\varphi_1-M\varphi_2+aM\varphi_3+\mathcal{O}(M^2,a^2)~,    &\text{if } \vert{\varphi}\vert <\frac{\pi}{2}~;\\
\pi-\varphi_1+M\varphi_2-aM\varphi_3+\mathcal{O}(M^2,a^2)~,&\text{if } \vert {\varphi}\vert >\frac{\pi}{2}~,\nn
\end{cases} \label{eq:phitwosol}
\end{eqnarray}
where
\begin{eqnarray}
&&\varphi_1=\arcsin{(bu)}~,\nn\\
&&\varphi_2=\frac{1+v^2-b^2u^2v^2}{b^2\sqrt{1-b^2u^2}v^2}~,\nn\\
&&\varphi_3=\frac{2}{b^2\sqrt{1-b^2u^2}v}~.\nn
\end{eqnarray}
By the lensing setup, substituting $u_S=1/r_S$ and $u_R=1/r_R$ for $u$ respectively into Eq. \eqref{eq:phitwosol},  one obtains
\begin{eqnarray}
\label{SRKerr}
\varphi_S&=&\arcsin{(bu_S)}-\frac{1+v^2-b^2u_S^2v^2}{b^2\sqrt{1-b^2u_S^2}v^2}M+\frac{2a M}{b^2\sqrt{1-b^2u_S^2}v}+\mathcal{O}(M^2,a^2)~,\nn\\
\varphi_R&=&\pi-\arcsin{(bu_R)}+\frac{1+v^2-b^2u_R^2v^2}{b^2\sqrt{1-b^2u_R^2}v^2}M-\frac{2a M}{b^2\sqrt{1-b^2u_R^2}v}+\mathcal{O}(M^2,a^2)~.
\end{eqnarray}
The advantage of Eq.~\eqref{SRKerr} is that we can express the finite-distance deflection angle using $u_S$ and $u_R$ as we will see in a moment.

Now, the surface integral of the Gaussian curvature can be carried out as
\begin{eqnarray}
\label{Gauss-def}
-\iint_{_{R}^{\infty}\Box_{S}^{\infty}}K_{\mathrm{K}} dS&=&-\int_{\varphi_S}^{\varphi_R}\int_{r(\varphi)}^{\infty}K_{\mathrm{K}}\sqrt{\det{\alpha^{\mathrm{K}}}}~dr d\varphi\nn\\
&=&\int_{\varphi_S}^{\varphi_R}\int_{u(\varphi)}^{0}u^{-2}K_{\mathrm{K}}\sqrt{\det\alpha^{\mathrm{K}}}~du d\varphi\nn\\
 &=&\int_{\varphi_S}^{\varphi_R}\int_0^{u(\varphi)}\bigg[\frac{(1+v^2)M}{v^2}+\frac{(6v^2+v^4-4)M^2u}{v^4} +\mathcal{O}(M^3,M^2a,a^2M,a^3)\bigg]~du d\varphi\nn\\
  &=&\frac{\left(1+v^2\right)\left(\sqrt{1-b^2u_R^2}+\sqrt{1-b^2u_S^2}\right)}{b v^2}M\nn\\
  &&+\frac{3\left(4+v^2\right)\left[\pi-\arcsin(b u_R)-\arcsin(b u_S)\right]}{4b^2v^2}M^2\nn\\
  &&+\frac{u_S\left[3v^2\left(4+v^2\right)+b^2\left(4-8v^2-3v^4\right)u_S^2\right]}{4bv^4\sqrt{1-b^2u_S^2}}M^2\nn\\
  &&+\frac{u_R\left[3v^2\left(4+v^2\right)+b^2\left(4-8v^2-3v^4\right)u_R^2\right]}{4bv^4\sqrt{1-b^2u_R^2}}M^2~\nn\\
  &&+\mathcal{O}(M^3,M^2a,a^2M,a^3)~,
\end{eqnarray}
where we have used Eqs.~\eqref{0th} and~\eqref{SRKerr}.

\subsection{Geodesic curvature}
Now we calculate the geodesic curvature of particle ray. Substituting Eq.~\eqref{KerrJMRE} into Eq.~\eqref{gc2}, the geodesic curvature of particle ray is
\begin{eqnarray}
\label{GEKerr}
 &&k_g^{\mathrm{K}}=-\frac{2\sqrt{1-v^2}}{mv^2}\frac{a M }{r^3}+\mathcal{O}(M^3,M^2a,a^2M,a^3)~,~~~~
\end{eqnarray}
where we have used Eq.~\eqref{EJ}.
We can obtain the transformation from Eq.~\eqref{Kerrline} as following
\begin{eqnarray}
\label{PTKerr}
 dl=\frac{m b v}{\sqrt{1-v^2}}\csc^2\varphi~ d\varphi+\mathcal{O}\left(M,a\right)~.~~~~
\end{eqnarray}
Then, one can get the part of deflection angle related to the path integral of geodesics using Eqs.~\eqref{GEKerr} and~\eqref{PTKerr}, as
\begin{eqnarray}
\label{Geodesic-def}
 \int_{S}^{R}k_g^{\mathrm{K}}dl&=&-\frac{2a M }{b^2v}\int_{\varphi_S}^{\varphi_R}\sin\varphi~d\varphi+\mathcal{O}(M^3,M^2a,a^2M,a^3)\nn\\
 &=&-\frac{2a M  \left(\sqrt{1-b^2u_R^2}+\sqrt{1-b^2u_S^2}\right)}{b^2v}+\mathcal{O}(M^3,M^2a,a^2M,a^3)~,~~~~~~~~
\end{eqnarray}
where we have also used Eqs.~\eqref{0th}, and~\eqref{SRKerr}. It should be noted that we have assumed that the particle orbit is prograde relative to the rotation of the Kerr BH and thus the sign of the right-hand side of Eq.~\eqref{Geodesic-def} changed if the particle ray is a retrograde orbit.
\subsection{Deflection angle}
By combining Eqs.~\eqref{Gauss-def} and~\eqref{Geodesic-def}, the finite-distance deflection angle of massive particles in Kerr spacetime can be written as
\begin{eqnarray}
\label{Kerr-1}
 \hat{\alpha}_{\mathrm{K}}&=&\frac{\left(1+v^2\right)\left(\sqrt{1-b^2u_R^2}+\sqrt{1-b^2u_S^2}\right)}{b v^2}M+\frac{3\left(4+v^2\right)\left[\pi-\arcsin(b u_R)-\arcsin(b u_S)\right]}{4b^2v^2}M^2\nn\\
  &&+\frac{u_S\left[3v^2\left(4+v^2\right)+b^2\left(4-8v^2-3v^4\right)u_S^2\right]}{4bv^4\sqrt{1-b^2u_S^2}}M^2+\frac{u_R\left[3v^2\left(4+v^2\right)+b^2\left(4-8v^2-3v^4\right)u_R^2\right]}{4bv^4\sqrt{1-b^2u_R^2}}M^2\nn\\
 &&\pm\frac{2a M  \left(\sqrt{1-b^2u_R^2}+\sqrt{1-b^2u_S^2}\right)}{b^2v}+\mathcal{O}\left(M^3,M^2a,a^2M,a^3\right)~,
\end{eqnarray}
where the positive and negative signs are for a retrograde and prograde particle rays, respectively. As expected, the transform from retrograde motion to prograde motion or vice versa is also equivalent to the sign change of the angular momentum $a$. It is also noticeable that if we exchange $u_R$ and $u_S$, the result is unchanged. This indeed is the consequence that the trajectory's radial coordinate is symmetric about the closest point. Note that for lightrays $v=1$, Eq.~\eqref{Kerr-1} leads to the result obtained by Ono \textit{et al} using the generalized optical metric method~\cite{OIA2017}. In addition, in the limit $u_S\rightarrow 0$ and $u_R\rightarrow 0$ ($i.e.$, $r_S\rightarrow \infty$ and $r_R\rightarrow \infty$ ), Eq.~\eqref{Kerr-1} reduces to the infinite-distance deflection angle of massive particles
\begin{eqnarray}
\label{Kerr-4}
 \hat{\alpha}_{\mathrm{K},\infty}&=&\frac{2\left(1+v^2\right)M}{b v^2}+\frac{3\pi \left(4+v^2\right)M^2}{4b^2 v^2}\pm\frac{4a M}{b^2v}+\mathcal{O}(M^3,M^2a,a^2M,a^3)~,~~~
\end{eqnarray}
which is consistent with the result reported in Refs.~\cite{He2017b,CGJ2019}. 

\begin{figure}[htp]
\begin{center}
\includegraphics[width=0.45\textwidth]{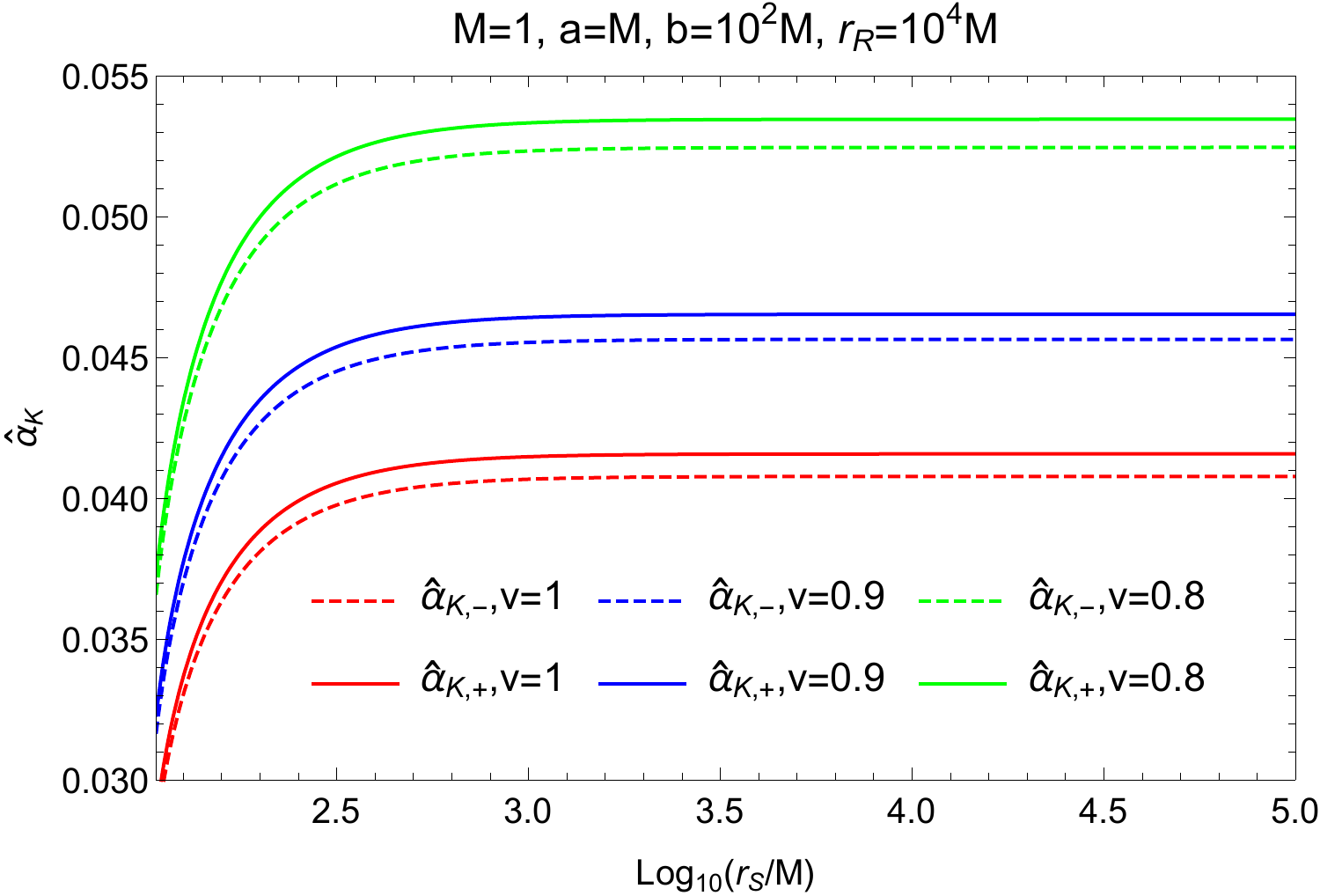} \hspace{0.05\textwidth}
\includegraphics[width=0.45\textwidth]{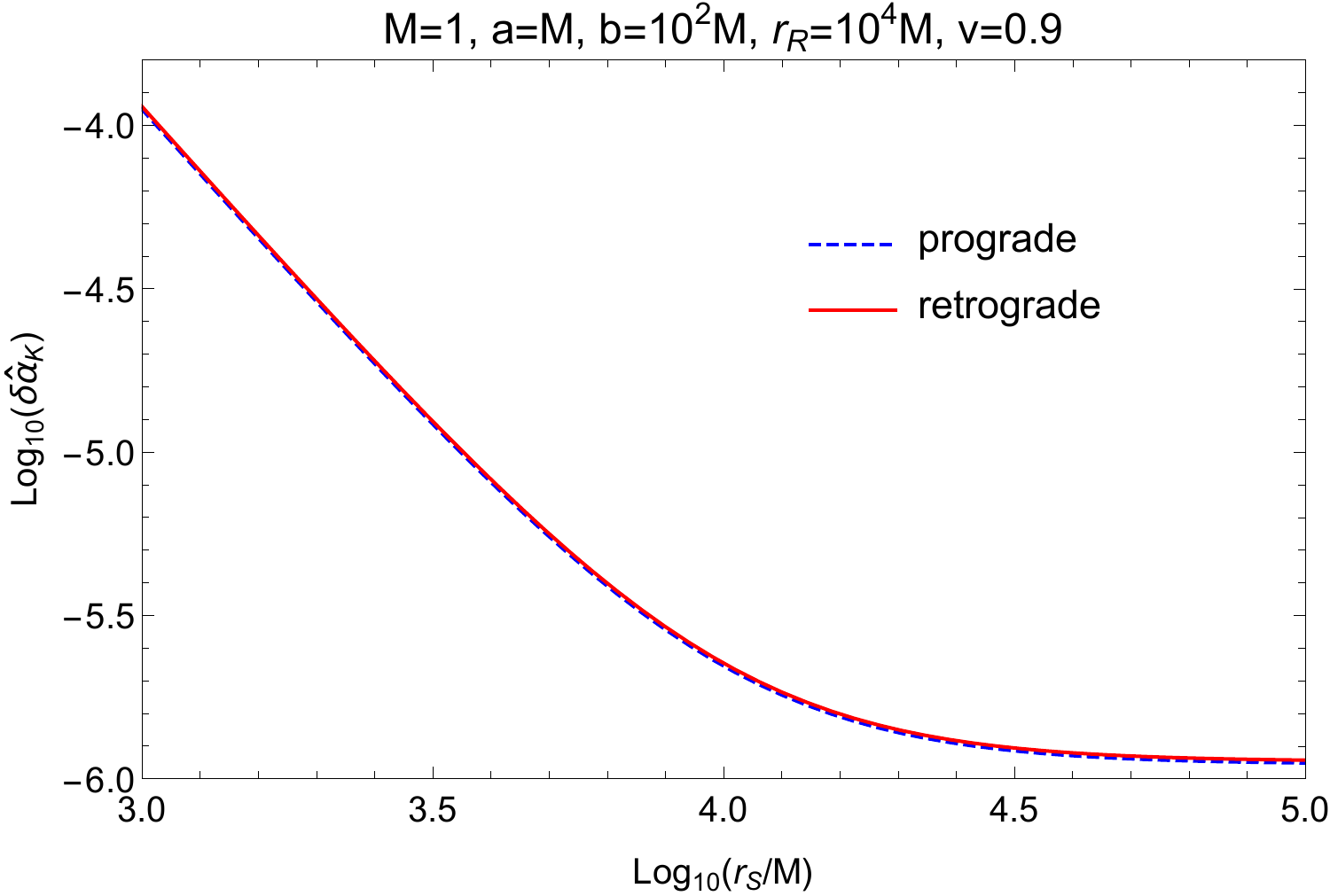}\\
(a) \hspace{0.5\textwidth}(b)
\end{center}
\caption{The finite-distance deflection angles of massless and massive particles in Kerr BH spacetime. (a) The deflection angle itself, Eq. \eqref{Kerr-1} as a function of $r_S$ for three velocities and the retrograde motion $(\hat{\alpha}_{\mathrm{K},+})$ and prograde motion $(\hat{\alpha}_{\mathrm{K},-})$; (b) The finite-distance correction, Eq. \eqref{Kerr-finite}, as a function of $r_S$ for a retrograde and prograde massive particle. }\label{Figure2}
\end{figure}

In Fig.~\ref{Figure2} (a), we show the deflection angle for a finite source distance. Here and henceforth we have set $M=1$ and measure other quantities with length dimension by $M$. Other parameter we choose are $b=10^2M, ~r_R=10^4M$ and for $a$ we scan it from 0 to $M$. It is seen that the deflection angle monotonically increases as $r_S$ increases from slightly larger than $b$ to $10r_R$ for all three velocities considered, $v/c=1,~0.9$ and 0.8. As $a$ changes from prograde motion (dashed line) to retrograde motion (solid line) for any fixed velocity, the deflection angle increases slightly. When the velocity decreases while holding other parameters, the deflection angle also increases. This is in agreement with what is found in Ref. \cite{Jia2016, Jia2019} and is expected because slower rays with same impact parameter tends to pass by the gravitational center more closely and therefore experience stronger bending to its trajectory, regardless whether the spacetime is rotating or not.

Comparing to deflection of lightrays in the Schwarzschild spacetime, from Eq. \eqref{Kerr-1} it is seen that there exist a few kinds of corrections for an ultra-relativistic ray originating from finite distance in a Kerr spacetime. The first is due to the presence of angular momentum of the Kerr spacetime (especially when $a$ is not large), while the second is due to the finite-distance effect and the third is due to the subluminal speed. 
Formula \eqref{Kerr-1} allows us to compare these three effects at once. Among these, the effect of a nonzero $a$ is most apparent, which is described by the second last term in this equation. For the finite-distance correction, expanding \eqref{Kerr-1} at $u_S=u_R=0$, we find 
\begin{eqnarray}
\label{eq:hakuexp}
\hat{\alpha}_{\mathrm{K}}=\hat{\alpha}_{\mathrm{K},\infty}+\delta\hat{\alpha}_{\mathrm{K},r}~,
\end{eqnarray}
where $\hat{\alpha}_{\mathrm{K},\infty}$ is given in Eq.~\eqref{Kerr-4} and $\delta\hat{\alpha}_{\mathrm{K},r}$ is
\begin{eqnarray}
\label{Kerr-finite}
 \delta\hat\alpha_{\mathrm{K},r}&\approx&\frac{\left(1+v^2\right)}{2 v^2}\left(\frac{M b}{r_R^2}+\frac{M b}{r_S^2}\right)+\frac{\left(-2+2v^2+v^4\right)}{2 v^4}\left(\frac{M^2 b}{r_R^3}+\frac{M^2 b}{r_S^3}\right)\pm\frac{1}{v}\left(\frac{a M}{r_R^2}+\frac{a M}{r_S^2}\right)~.
\end{eqnarray}
Equation \eqref{Kerr-finite} can be thought as the finite-distance correction for general velocity $v$ and angular momentum $a$. It is clear in this equation that for $a\leq M\ll b$ and velocity that is not too small, the first term will dominate the third one which involves angular momentum. Therefore this finite-distance effect is hardly affected by the spacetime rotation. This is verified in Fig. \ref{Figure2} (b), in which we plot $\Delta \hat{\alpha}_{\mathrm{K,r}}$ for $a=M,~b=10^2M, ~r_R=10^4M$ and $v=0.9c$, it is seen that this quantity changes about two orders when $r_S$ goes from $10^3M$ to $10^5M$ while the two curves for retrograde and prograde motions (or equivalently $\pm a$) almost overlap. Moreover, one can also see from Eq. \eqref{Kerr-finite} that for the above parameters, the finite-distance effect decreases monotonically as $r_S$ increases, which is again expected. 

Further expanding Eq. \eqref{eq:hakuexp} around $v/c=1$ and $a=0$, to the first nontrivial orders of $u_R$ or $u_S$, $1-v$ and $a$, and to the first order of $M$, we obtain 
\begin{align}
\label{eq:hakuexpvaexp}
\hat{\alpha}_{\mathrm{K}}=\hat{\alpha}_{\mathrm{S},\infty,c}+\Delta\hat{\alpha}_a+\Delta\hat{\alpha}_v+\Delta\hat{\alpha}_r~,
\end{align}
where 
\begin{eqnarray}
&&\hat{\alpha}_{\mathrm{S},\infty,c}=\frac{4M}{b},\\
&&\Delta\hat{\alpha}_a=\frac{4M}{b^2}a,\\
&&\Delta\hat{\alpha}_v=\frac{4M}{b}(1-v),\\
&&\Delta\hat{\alpha}_r=Mb\left(\frac{1}{r_R^2}+\frac{1}{r_S^2}\right),
\end{eqnarray}
We wish to compare the sizes of these three corrections, and if possible in which part of the relevant parameter space spanned by $(b,~v,~r_S,~r_R,~a)$ any of them will dominate others. 
To do this, we approximate the impact parameter $b$ by the geometric relation $b\approx \theta r_R$ in GL, where $\theta$ is the apparent angle of the GL images, and then study three ratios 
\begin{align}
    &\displaystyle \frac{\Delta\hat{\alpha}_a}{\Delta\hat{\alpha}_r}= \frac{a/M}{\frac{\theta^3 r_R/M}{4}\left(1+\frac{r_R^2}{r_S^2}\right)}, \label{eq:arratio}\\
    &\displaystyle \frac{\Delta\hat{\alpha}_v}{\Delta\hat{\alpha}_r}= \frac{1-v}{\frac{\theta^2}{4}\left(1+\frac{r_R^2}{r_S^2}\right)},\label{eq:vrratio}\\
    &\displaystyle \frac{\Delta\hat{\alpha}_a}{\Delta\hat{\alpha}_v}= \frac{a/M}{(1-v)\theta r_R/M}. \label{eq:avratio}
\end{align}
Clearly, if the angular momentum $a$ (or the velocity difference $(1-v)$) is larger than the denominator of Eq. \eqref{eq:arratio} (or Eq. \eqref{eq:vrratio}), then the correction of spacetime rotation (or velocity) will be larger than that of finite-distance. Otherwise, the opposite happens. If Eq. \eqref{eq:avratio} is larger than one, the effect of nonzero $a$ dominates the effect of velocity. 

The typical values of these parameters for the GL by galaxies or galaxy clusters, and for the microlensings in the star-planet systems, can be obtained respectively from two large data sets  \cite{gl:extra,gl:micro}. 
The apparent angle usually ranges between $\theta_{g,l}=0.34$ [as] to $\theta_{g,u}=22.5$ [as] for GL by extragalactic objects, and from 0.09 [mas] to 1.45 [mas] for microlensings. The lens distance $r_R$ can vary from redshift $z=0.04$ to $z=1.01$ for the former case and from 380 [pc] to 8800 [pc] for the latter case. While the $r_R/r_S$ ratio in both cases range from $0.1$ to $10$. Using these data, therefore we can attempt to study the ratios in Eqs. \eqref{eq:arratio} to \eqref{eq:avratio}. However for the numerator $a$ in Eq. \eqref{eq:arratio}, since we usually do not know its absolute value but its ratio with respect to the lens mass, i.e., $a/M$, the mass $M$ of the lens should also be known. Although the lens mass in the microlensing cases are easily obtained (e.g., in \cite{gl:micro} about 36 lensings have $M$ estimated), the masses of the galaxy and galaxy clusters lenses are usually not provided by data (e.g. \cite{gl:extra}). Therefore we will not study GLs of galaxy and galaxy cluster, but that of some supermassive BHs in centers of galaxies. We use the BH data, including their masses and distances provided in Ref. \cite{Virbhadra:2008ws, Saglia:2016sinfoni}, and further assume that the GL they might cause will also yield apparent angles $\theta$ that is in the range of $(\theta_{g,l},~\theta_{g,u})$ and $r_R/r_S$ ratio between $0.1$ to $10$. These assumptions are expected to be reasonable for GLs by BHs. 

\begin{center}
\begin{figure}[htp]
\includegraphics[width=0.3\textwidth]{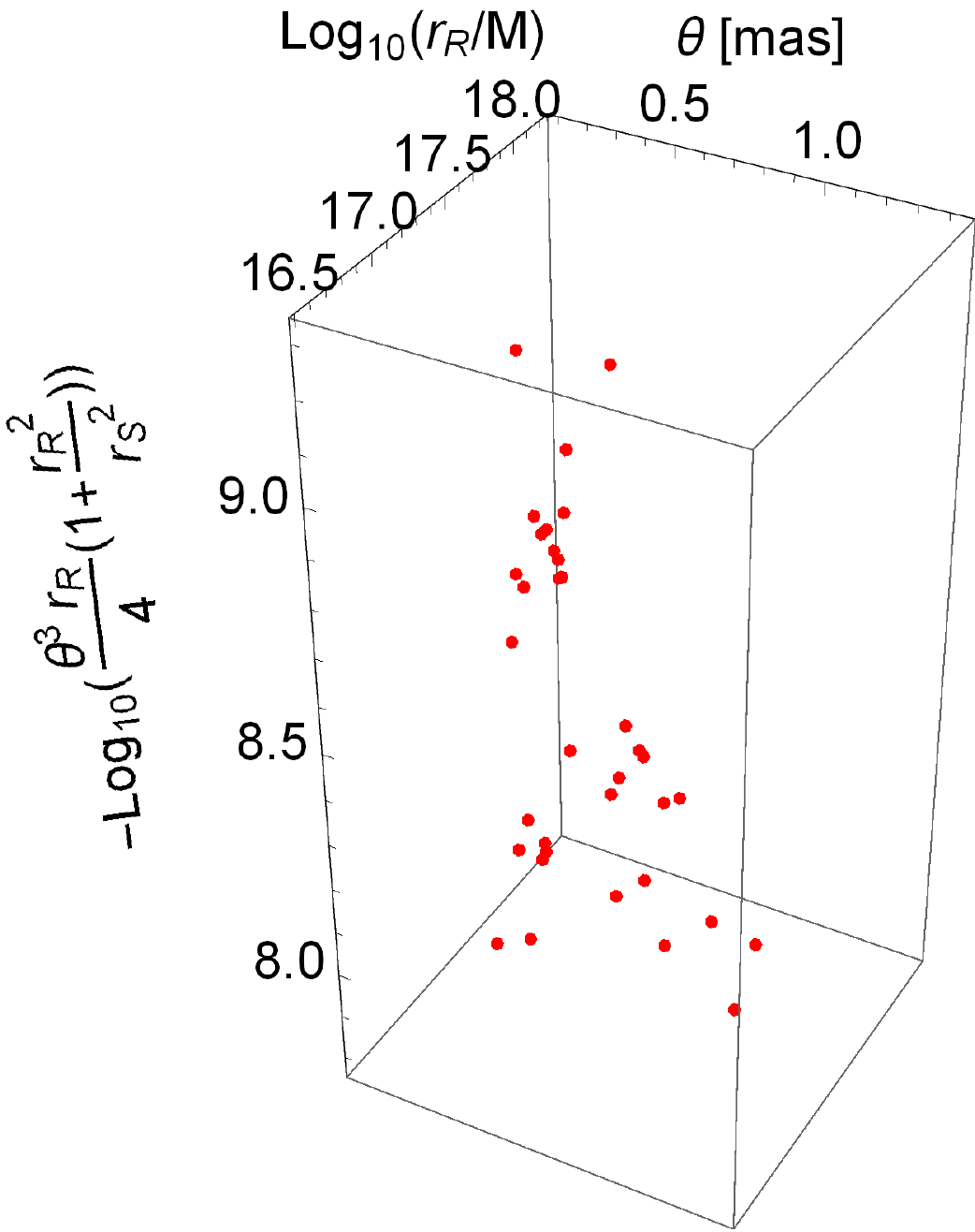}\hspace{0.04\textwidth}
\includegraphics[width=0.3\textwidth]{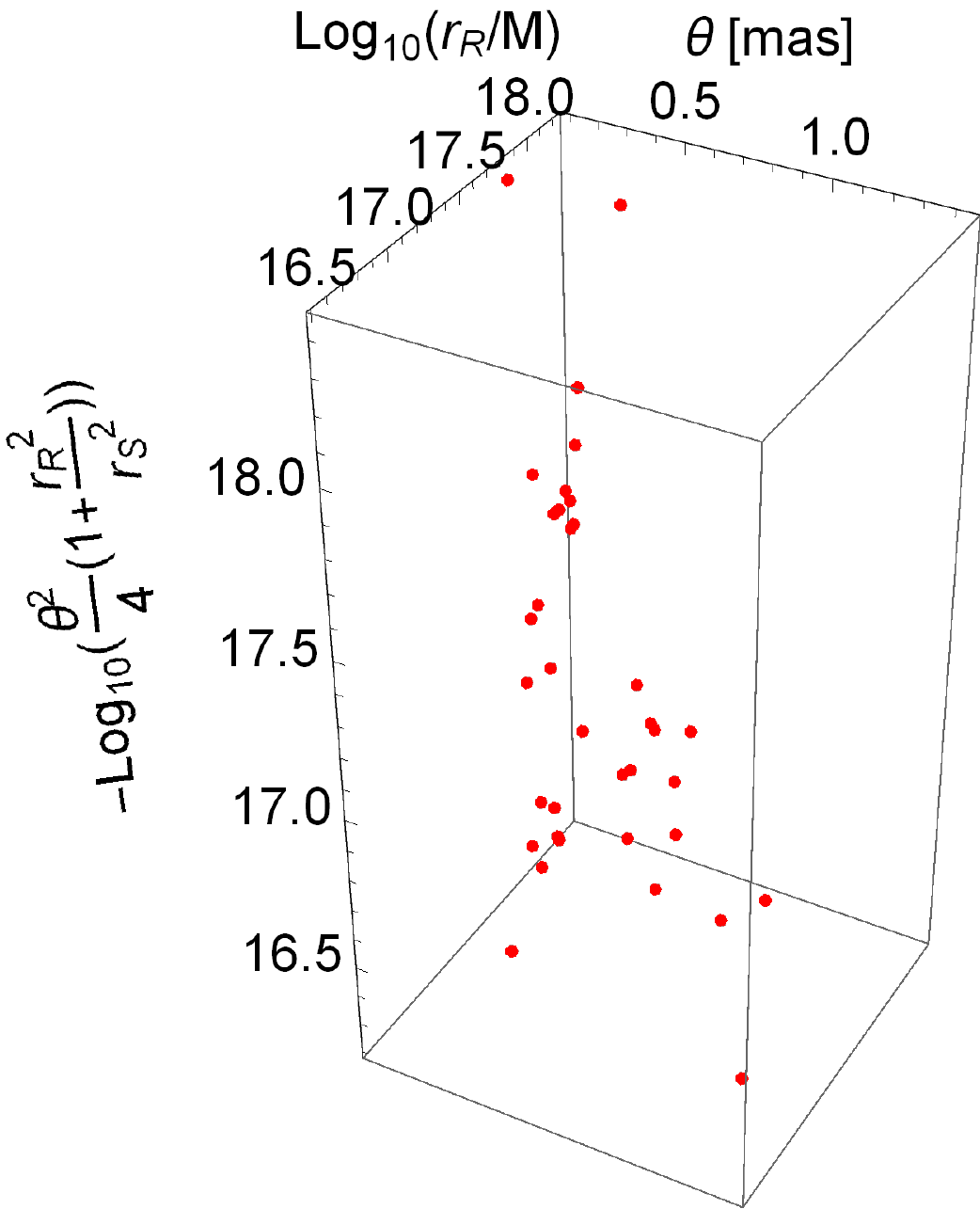}\hspace{0.04\textwidth}
\includegraphics[width=0.265\textwidth]{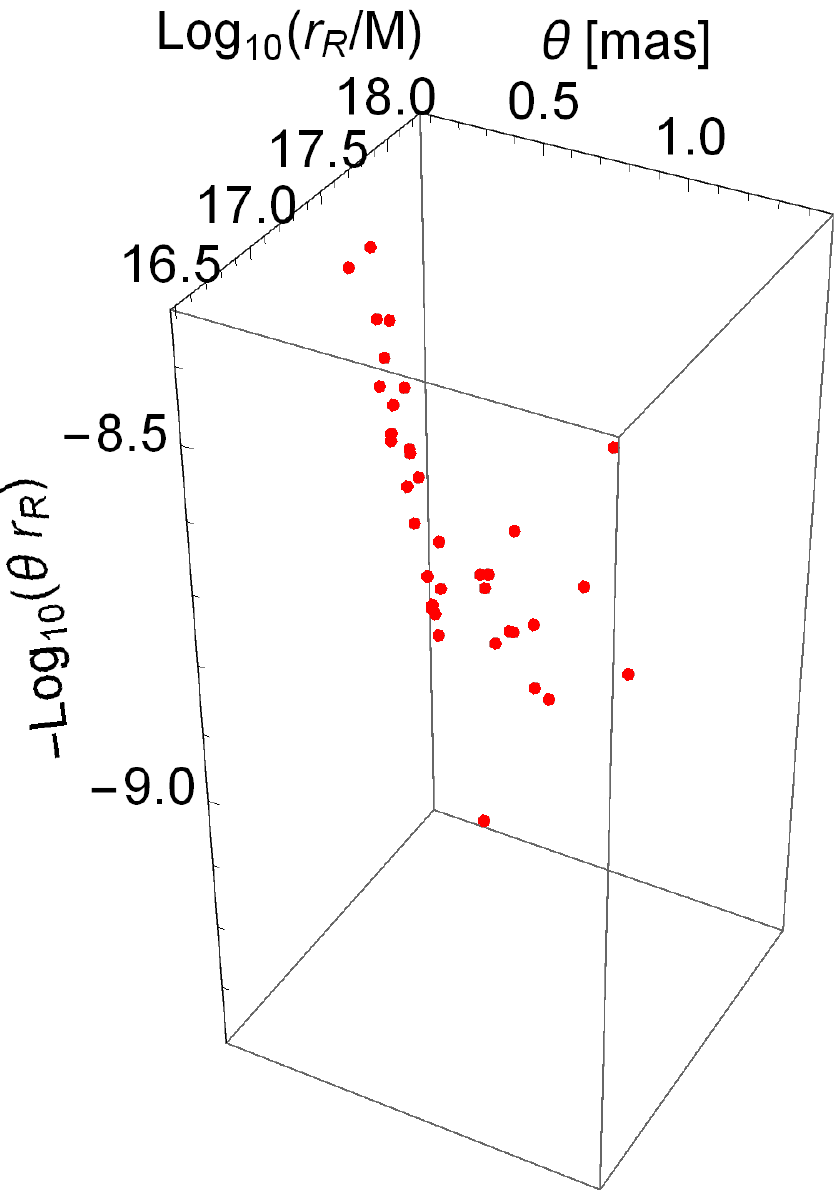}\\
(a)\hspace{0.3\textwidth}(b)\hspace{0.3\textwidth}(c)\\
\includegraphics[width=0.3\textwidth]{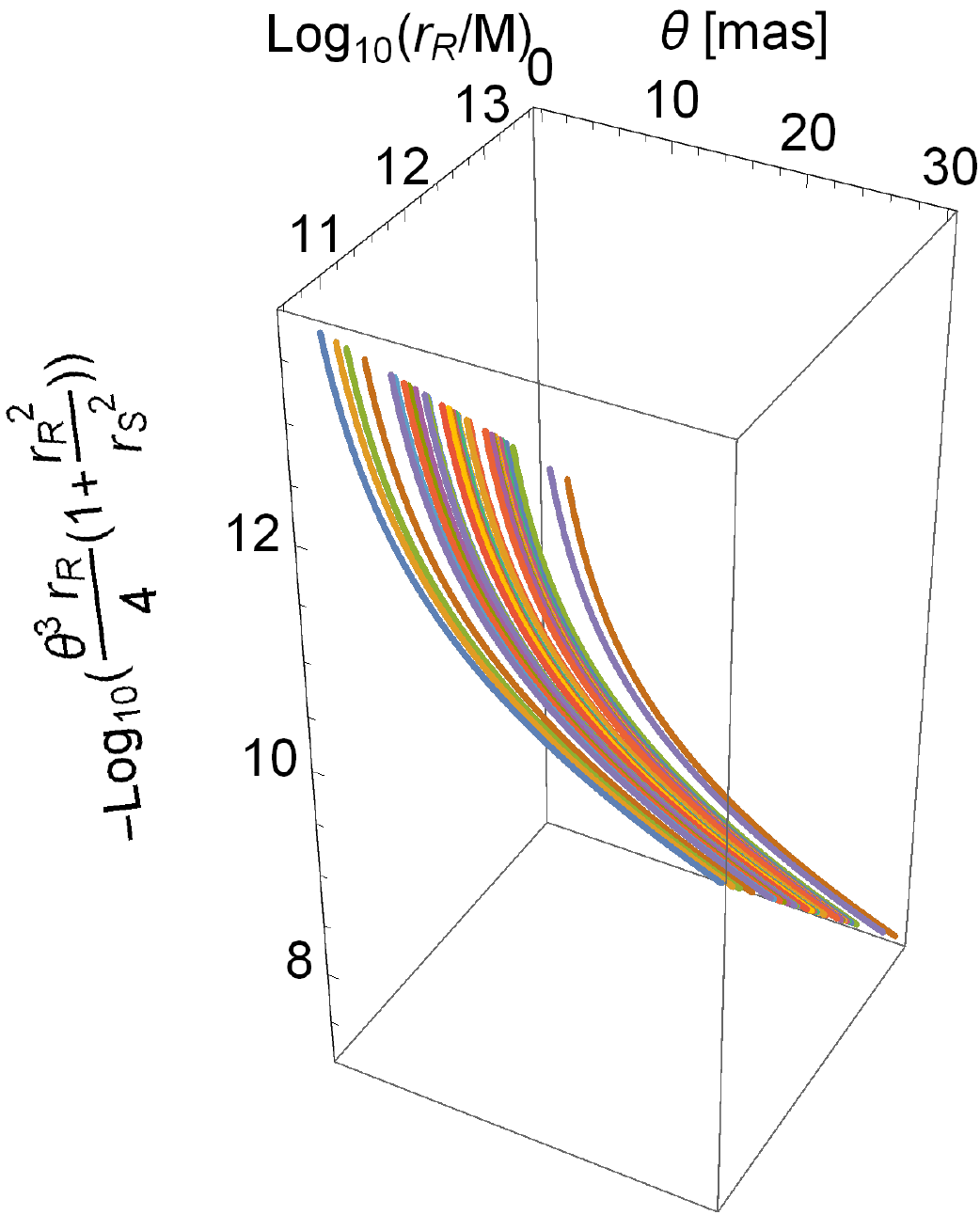}\hspace{0.04\textwidth}
\includegraphics[width=0.3\textwidth]{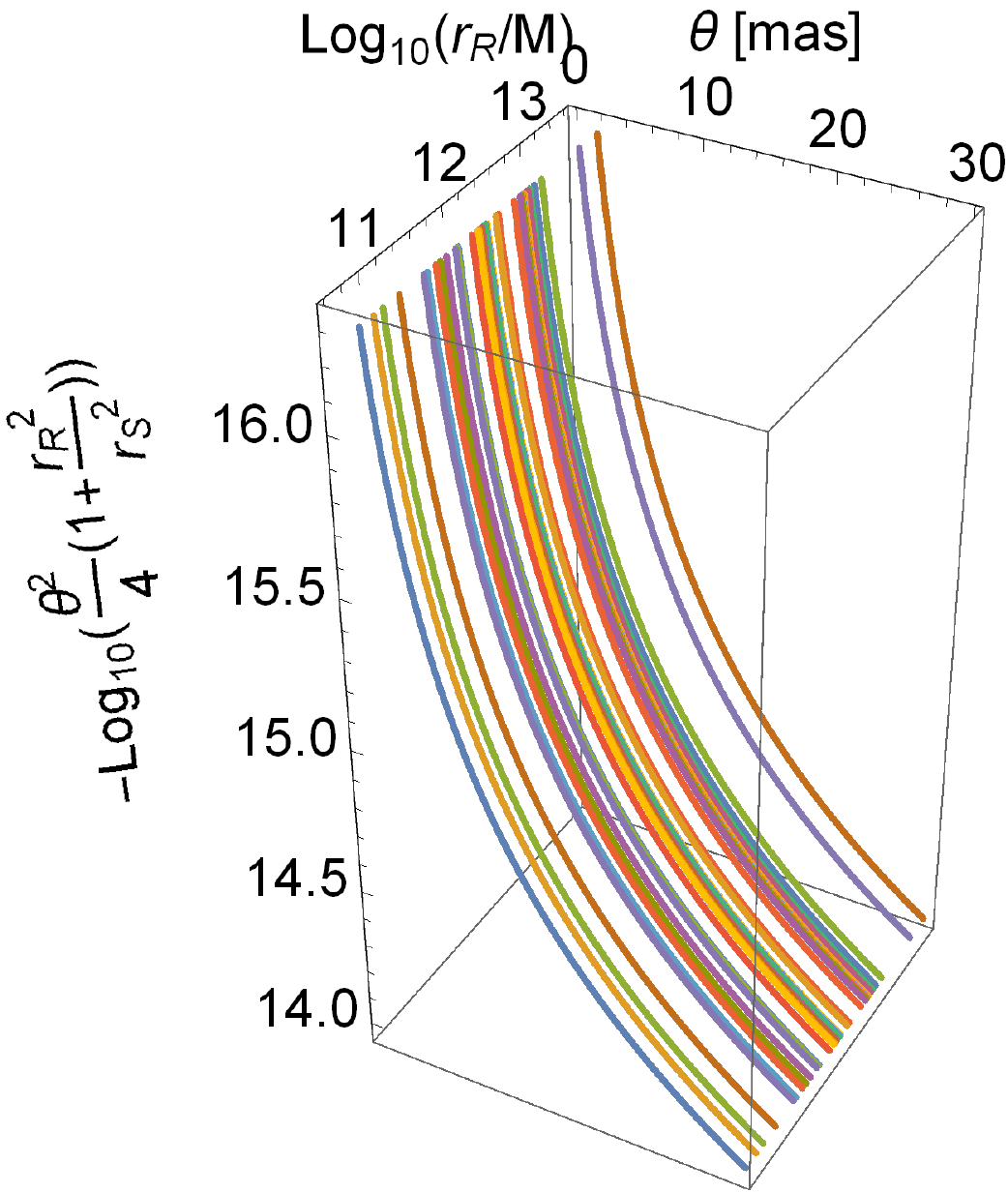}\hspace{0.04\textwidth}
\includegraphics[width=0.25\textwidth]{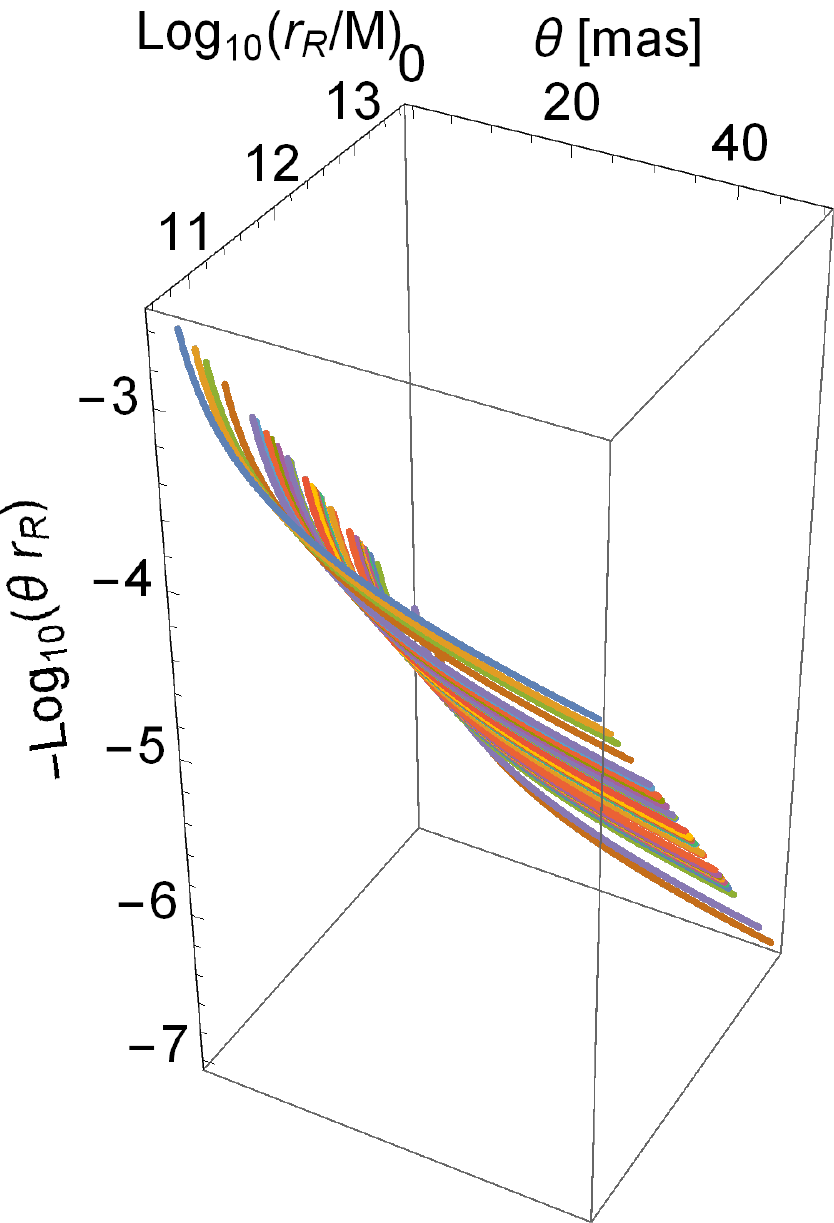}\\
(d)\hspace{0.3\textwidth}(e)\hspace{0.3\textwidth}(f)
\caption{Coefficients of (a) $a/M$ in Eq. \eqref{eq:arratio}, (b) $(1-v)$ in Eq. \eqref{eq:vrratio} and (c) $a/[M(1-v)]$ in Eq. \eqref{eq:avratio}, for the microlensing case. The same coefficients for the BH lensing are in (d)-(f). $r_R/r_S=1$ was used. Data are from Ref. \cite{gl:micro}.} \label{fig:ratiokerr}
\end{figure}
\end{center}

In Fig. \ref{fig:ratiokerr} (a), (b) and (c) respectively, we plot
the coefficient of $a/M$ in Eq. \eqref{eq:arratio}, the coefficient of $(1-v)$ in Eq. \eqref{eq:vrratio} and that of $a/[M(1-v)]$ in Eq. \eqref{eq:avratio} for the microlensing case. The corresponding plots for the BH lensing case are plotted in Fig. \ref{fig:ratiokerr} (d)-(f). 
It is seen from Fig. \ref{fig:ratiokerr} (a), (d) that the typical value of the coefficient of $a/M$ in Eq. \eqref{eq:arratio} is in the range of $10^{7.5}$ to $10^{9.5}$ for the microlensing, and $10^7$ to $10^{16}$ for supermassive BH lensing, which is slightly wider due to the larger variation of $\theta$. Since $a/M$ in typical Kerr BHs are much larger than $10^{-7}$ (e.g. all Kerr BH in detected GW events), then this suggests that practically for all Kerr BHs the effect of BH spin to the deflection angle is much larger than that of a finite distance of the source or receiver. 

From Fig. \ref{fig:ratiokerr} (b), (e), one then see that the coefficient of $(1-v/c)$ in Eq. \eqref{eq:vrratio} is in the range of $10^{16}$ to $10^{18.5}$ for microlensing and $10^{14}$ to $10^{18}$ for supermassive BH lensings. This implies that when $(1-v/c)>10^{-14}$ then the effect of velocity will most likely be larger than that of finite distance in both kinds of GLs. If $10^{-18.5}<(1-v/c)<10^{-16}$ in the microlensing case or $10^{-18}<(1-v/c)10^{-14}$ in the supermassive BH lensing case, then there is still chance that effect of velocity is larger than finite radius. Otherwise, the effect of finite source or receiver radius will be larger.  For lensed supernova neutrinos, using their typical energy of the order of $\mathcal{O}(10)$ [MeV], one can work out that the above velocity ranges require the lensed neutrino mass-eigenstate to have a mass $m_i$ larger than 1.4 [eV] to guarantee  a larger effect than the finite distance. Otherwise, if  $0.008\text{~[eV]}<m_i<1.4\text{~[eV]}$, then the relative size of the velocity effect and finite radius will depend on the specific lens parameters. If $m_i<0.008$ [eV], then typical finite size effect will be larger than the effect of velocity. Note that all these above mass ranges are still allowed by current neutrino mass constraints, i.e., $\Delta m_{21}^2=7.53\times 10^{-5}$ [eV$^2$] and $|\Delta m_{23}^2|\approx 2.5\times 10^{-3}$ [eV$^2$] \cite{Tanabashi:2018oca}. For GWs, their previously measured velocity is constrained to $(1-3\times 10^{-15})c<v<(1+10^{-16})c$ \cite{Monitor:2017mdv}. Therefore for both microlensings and GL by galaxies, depending on the exact GW velocity and the lens parameters, the effect of velocity to the deflection angle might be larger or smaller than that of the finite distance of the source or receiver.

\section{Teo wormholes}\label{TEO}

The Teo wormhole metric describes a stationary, axisymmetric and
asymptotically flat rotating wormhole spacetime given by~\cite{Teo1998}
\begin{eqnarray}
\label{0R-ABC}
 ds^2&=&N^2 dt^2-\frac{dr^2}{(1-\frac{b_0}{r})}-r^2 H^2\left[d\theta^2+\sin^2 \theta(d\varphi-wdt^2)\right]~,
\end{eqnarray}
where
\begin{eqnarray}
&&N=H=1+\frac{\lambda(4a_0 \cos\theta)^2}{r}~,\label{eq:nhdef}\\
&&\omega=\frac{2a_0}{r^3}~.
\end{eqnarray}
Here $a_0$ is the total angular momentum of the wormhole, $b_0$ represents the shape function with the condition $r\geq b_0$ and $\lambda$ is a constant.
Using Eq. \eqref{spatial metric} and \eqref{Randers-Finsler} for the metric \eqref{0R-ABC}, one can deduce respectively
\begin{eqnarray}
 \gamma _{ij}^{\mathrm{T}}{dx}^idx^j&=&\frac{\text{dr}^2}{1-\frac{b_0}{r}}+H^2r^2\text{d}\theta ^2+\bigg( H^2r^2\sin ^2\theta+\frac{H^4r^4\omega ^2\sin ^4\theta}{N^2-H^2r^2\omega ^2\sin ^2\theta } \bigg) \text{d}\varphi ^2,
\end{eqnarray}
and the Teo JMRF metric
\begin{eqnarray}
\label{GJTeo}
 \alpha_{ij}^{\mathrm{T}}dx^i dx^j&=&\left( \frac{\mathcal{E}^2}{ N^2-H^2r^2\omega ^2 \sin^2\theta}-m^2 \right)\nn\\
 &&\times\bigg[ \frac{{dr}^2}{1-\frac{b_0}{r}}+H^2r^2d\theta^2+\bigg(H^2r^2\sin^2\theta+\frac{H^4r^4\omega^2\sin^4\theta}{N^2-H^2r^2w^2\sin^2\theta}\bigg)d\varphi^2\bigg]~,\nn\\
\beta_i^{\mathrm{T}} dx^i&=&-\frac{\mathcal{E}H^2r^2\omega \sin ^2\theta \,\,\text{d}\varphi}{ N^2-H^2r^2\omega ^2\sin ^2\theta }~.
\end{eqnarray}
For $m=0$ and $\mathcal{E}=1$, the Teo-JMRF metric reduces to the Teo-ORF metric~\cite{Jus-183}.

\subsection{Gaussian curvature}
In the equatorial plane $\theta=\pi/2$, from Eq. \eqref{eq:nhdef} thus we have $H=N=1$ and the constant $\lambda$ in
the metric will not contribute. The generalized Teo Jacobi metric induced in the equatorial plane becomes
\begin{eqnarray}
\label{Teo-line}
dl^2=\alpha_{ij}^{\mathrm{T}}dx^idx^j=m^2\left[ \frac{1}{ \left(1-r^2\omega ^2 \right) \left(1-v^2\right)}-1 \right]\bigg[ \frac{{dr}^2}{1-\frac{b_0}{r}}+\left(r^2 +\frac{r^4\omega^2}{1-r^2w^2}\right)d\varphi^2\bigg]~,
\end{eqnarray}
where we have used Eq.~\eqref{EJ}.
Then, the corresponding Gaussian curvature is
\begin{eqnarray}
&& K_{\mathrm{T}}=-\frac{\left(1-v^2\right)b_0}{2m^2r^3v^2}+\mathcal{O}(b_0^3,a_0b_0,a_0^2)~.~~~
\end{eqnarray}
For metric~\eqref{0R-ABC}, the particle orbit equation~\eqref{Sorbit} can be iteratively solved as
\begin{eqnarray}
\label{orbitTeo}
&& u(\varphi)=\frac{\sin \varphi}{b}+\frac{\cos^2\varphi}{2b^2}b_0-\frac{2a_0}{b^3}+\mathcal{O}(b_0^2,a_0b_0,a_0^2)~.~~
\end{eqnarray}
Similar to the Kerr case, one can iteratively inverse the function $u(\varphi)$ and solve 
\begin{eqnarray}
\label{SRTeo}
\varphi_S&=&\arcsin{(bu_S)}-\frac{\sqrt{1-b^2u_S^2}}{2b}b_0+\frac{2a_0}{b^2\sqrt{1-b^2u_S^2}}+\mathcal{O}(b_0^2,a_0b_0,a_0^2)~,~~~\nn\\
\varphi_R&=&\pi-\arcsin{(bu_R)}+\frac{\sqrt{1-b^2u_R^2}}{2b}b_0-\frac{2a_0}{b^2\sqrt{1-b^2u_R^2}}+\mathcal{O}(b_0^2,a_0b_0,a_0^2)~.
\end{eqnarray}
The deflection angle related to the surface integral of Gaussian curvature is
\begin{eqnarray}
\label{Gauss-Teo}
-\iint_{_{R}^{\infty}\Box_{S}^{\infty}}K_{\mathrm{T}} dS&=&-\int_{\varphi_S}^{\phi_R}\int_{r(\varphi)}^{\infty}K_{\mathrm{T}}\sqrt{\det{\alpha^{\mathrm{T}}}}~dr d\varphi\label{eq:ktappear}\\
&=&\int_{\varphi_S}^{\varphi_R}\int_{u(\varphi)}^{0}u^{-2}K_{\mathrm{T}}\sqrt{\det\alpha^{\mathrm{T}}}~du d\varphi\nn\\
 &=&\int_{\varphi_S}^{\varphi_R}\int_0^{u(\varphi)}\left(\frac{b_0}{2}+\frac{b_0^2u}{4}+\mathcal{O}(b_0^3,a_0b_0,a_0^2)\right)~du d\varphi\nn\\
  &=&\frac{b_0\left(\sqrt{1-b^2u_R^2}+\sqrt{1-b^2u_S^2}\right)}{2b}+\frac{3\left[\pi-\arcsin(b u_R)-\arcsin(b u_S)\right]}{16b^2}b_0^2\nn\\
 &&-\frac{u_R\sqrt{1-b^2u_R^2}+u_S\sqrt{1-b^2u_S^2}}{16b}b_0^2+\mathcal{O}(b_0^3,a_0b_0,a_0^2)~,~~~~
\end{eqnarray}
where we have used Eqs.~\eqref{orbitTeo} and~\eqref{SRTeo}.
Note that in Eq. \eqref{eq:ktappear} although the Gaussian curvature $K_{\mathrm{T}}$ is dependent on particle's velocity $v$, $K_{\mathrm{T}}\sqrt{\det\alpha^{\mathrm{T}}}$ is not. Thus, the result of the surface integral of Gaussian curvature to the above orders is independent of particle velocity.

\subsection{Geodesic curvature}
Considering the three-dimensional generalized Teo Jacobi metric in Eq.~\eqref{GJTeo}, the geodesic curvature of particle ray can be calculated using Eqs.~\eqref{gc2} and ~\eqref{EJ} and the result is given by
\begin{eqnarray}
 &&k_g^{\mathrm{T}}=-\frac{2a_0 \sqrt{1-v^2}}{mv^2r^3}+\mathcal{O}\left(b_0^3,a_0b_0\right)~.
\end{eqnarray}
With the help of Eq.~\eqref{orbitTeo}, the parameter transformation can be obtained using Eq.~\eqref{Teo-line} as 
\begin{eqnarray}
 &&dl=\frac{b m v}{\sqrt{1-v^2}}\csc^2 \varphi~ d\varphi+\mathcal{O}\left(b_0,a_0\right)~.~~~~
\end{eqnarray}
Then, the path integral of geodesic curvature along the particle ray can be obtained after using Eqs.~\eqref{orbitTeo} and~\eqref{SRTeo} as
\begin{eqnarray}
\label{Geodesic-Teo}
\int_{S}^{R}k_g^{\mathrm{T}}dl&=&-\frac{2a_0}{b^2v}\int_{\varphi_S}^{\varphi_R}\sin\varphi~d\varphi+\mathcal{O}(b_0^3,a_0b_0,a_0^2)\nn\\
&=&-\frac{2a_0 \left(\sqrt{1-b^2u_R^2}+\sqrt{1-b^2u_S^2}\right)}{b^2v}+\mathcal{O}(b_0^3,a_0b_0,a_0^2)~.~~~~
\end{eqnarray}
Unlike the surface integral of Gaussian curvature in Eq.~\eqref{Gauss-Teo}, the result in Eq.~\eqref{Geodesic-Teo} is affected by the particle velocity and thus this term is different from the light deflection.

\subsection{Deflection angle}
By combining Eqs.~\eqref{Gauss-Teo} and~\eqref{Geodesic-Teo}, we get the deflection angle of massive particle for the receiver and source at finite distance from Teo wormhole lens as the following
\begin{eqnarray}
\label{Teo-1}
 \hat{\alpha}_{\mathrm{T}}&=&\frac{\left(\sqrt{1-b^2u_R^2}+\sqrt{1-b^2u_S^2}\right)}{2b}b_0+\frac{3\left[\pi-\arcsin(b u_R)-\arcsin(b u_S)\right]}{16b^2}b_0^2\nn\\
 &&-\frac{u_R\sqrt{1-b^2u_R^2}+u_S\sqrt{1-b^2u_S^2}}{16b}b_0^2\pm\frac{2a_0 \left(\sqrt{1-b^2u_R^2}+\sqrt{1-b^2u_S^2}\right)}{b^2v}+\mathcal{O}(b_0^3,a_0b_0,a_0^2)~,~~~~
\end{eqnarray}
where the positive and negative signs are for retrograde and prograde particle rays, respectively. Note that the terms involving $b_0^1$ and $b_0^2$ are not affected by the velocity of the particles, but the term containing $a_0^1$ is. It is worthwhile to
mention that unlike the divergence of $\hat{\alpha}_{\mathrm{K}}$ in the Kerr sapcetime, the finite-distance deflection angle $\hat{\alpha}_{\mathrm{T}}$ is finite for any one or both of the limits $r_S\rightarrow b$ and $r_R\rightarrow b$.
This is understandable from the fact that there is no event horizon for the Teo wormhole spacetime and therefore the particle rays do not tend to be bent infinitely even when their $r_S$ and $r_R$ appraoch their limits. 
More, it is also noticeable that the deflection angle is unchanged when switching $r_R$ and $r_S$, and there is no deflection when $r_S=r_R= b$. The result \eqref{Teo-1} in leading order agrees with the finite-distance deflection angle of light ($v=1$) obtained by Ono \textit{et. al} in Ref.~\cite{OIA2018}.

\begin{figure}[htp]
\begin{center}
\includegraphics[width=0.45\textwidth]{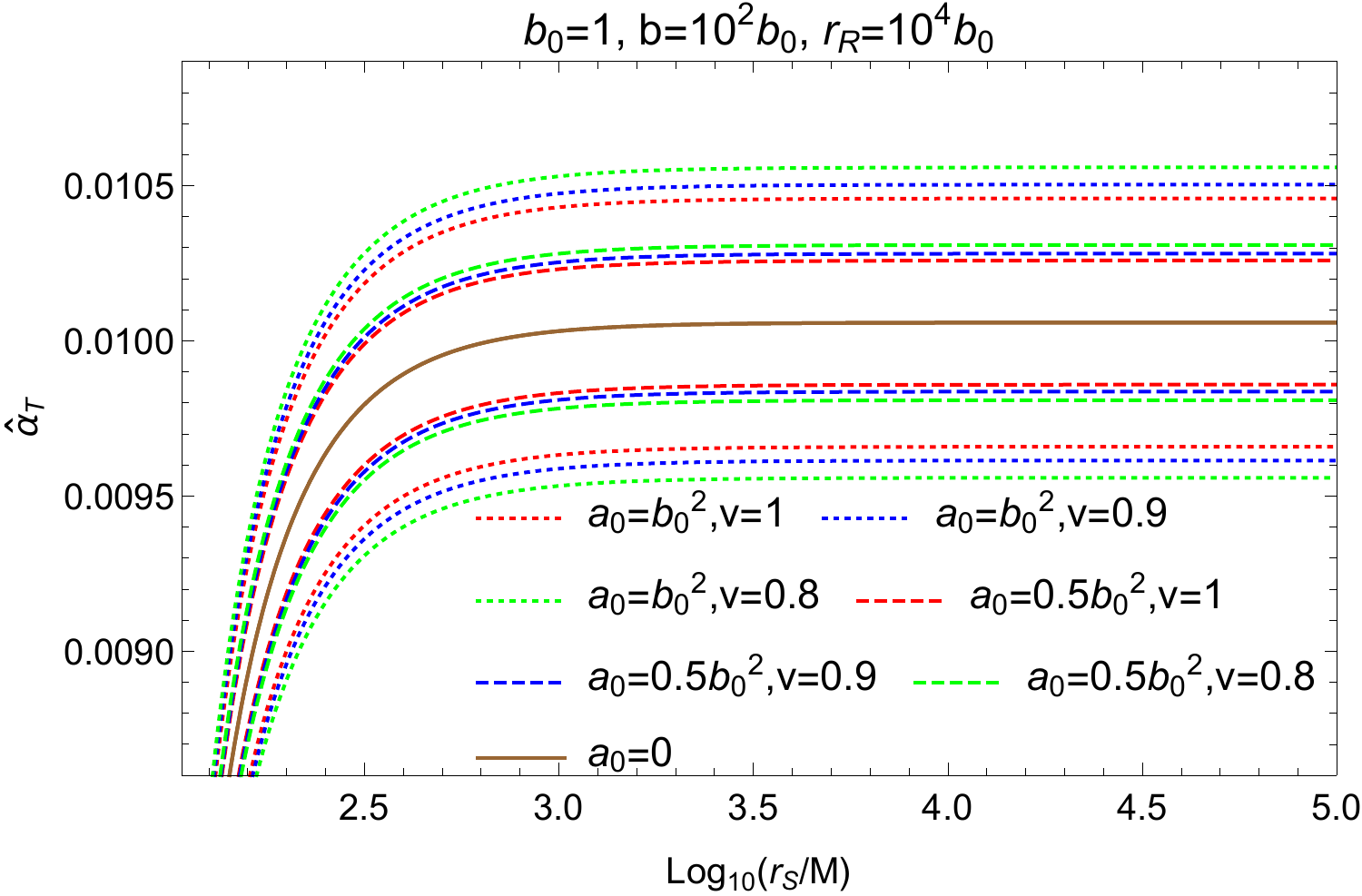} \hspace{0.05\textwidth}
\includegraphics[width=0.45\textwidth]{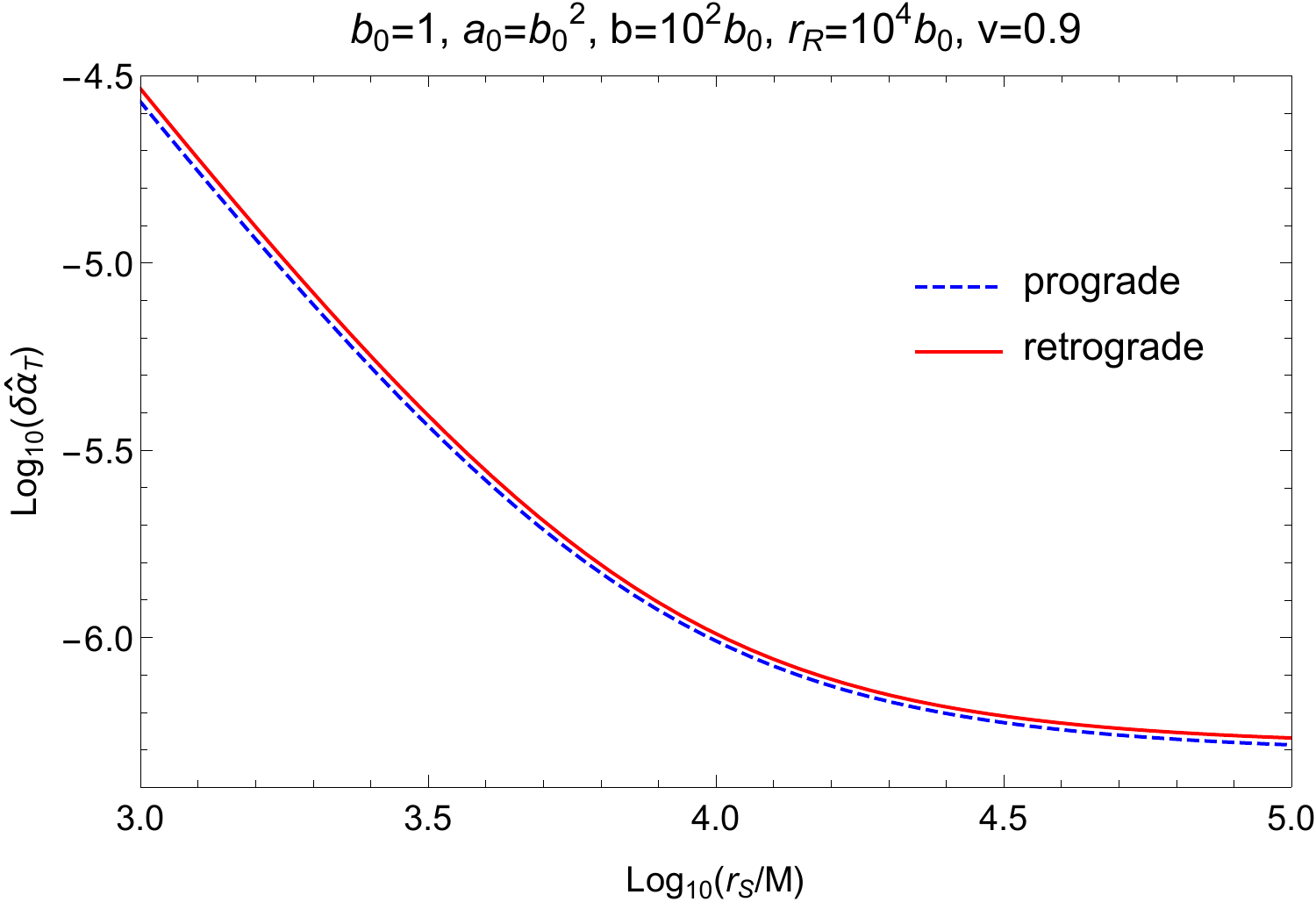}
\\
(a) \hspace{0.5\textwidth}(b)
\end{center}
\caption{The finite-distance deflection angle of massless and massive particles in Teo wormhole spacetime. (a) The deflection angle, Eq. \eqref{Teo-1}. The lines above the $a_0=0$ lines are for retrograde motion and the lines below are for retrograde motions.  (b) The finite distance correction, Eq. \eqref{Teo-finite}. }\label{Figure4}
\end{figure}

In Fig.~\ref{Figure4} (a), we plot the deflection angle \eqref{Teo-1} by fixing $b_0=1$ and measuring other quantities with length dimension by $b_0$. Similar to Fig. \ref{Figure2}, we choose $b=10^2b_0,~r_R=10^4b_0$ and three representative values of the angular momentum $a_0=0,~0.5b_0^2, ~b_0^2$ for both retrograde and prograde motions and three velocities $v/c=1,~0.9,~0.8$. It is seen that the total deflection angle increases as the receiver distance increases for all values of $a_0$ and $v$. As $a_0$ varies from retrograde with angular momentum $b_0^2$ to prograde with same size, the deflection angle decreases monotonically for all velocities and $r_R$. However, unlike the Kerr case where velocity's decrease always increases the deflection angle regardless in the retrograde or prograde cases, here as velocity decreases the prograde deflection angle decreases while the retrograde deflection angle increases. This is understand from the second last term of Eq. \eqref{Teo-1} that the only dependence of the deflection angle on $a_0$ and $v$ are correlated in a ratio form. This special from also determines completely how and to what extent the change of $a_0$ and $v$ will affect the deflection angle. 

When $r_R$ and $r_S$ are large but still finite, we can expand Eq. \eqref{Teo-1} to the first nontrivial orders of $1/r_R$ and $1/r_S$, to obtain 
\begin{align}
    \hat{\alpha}_T=\hat{\alpha}_{\mathrm{T},\infty}+\delta\hat\alpha_{\mathrm{T},r} \label{Teo-expr}
\end{align}
where the infinite distance deflection angle $\hat{\alpha}_{\mathrm{T},\infty}$ and its correction are respectively
\begin{eqnarray}
&&\hat{\alpha}_{\mathrm{T},\infty}=\frac{b_0}{b}+\frac{3\pi b_0^2}{16b^2}\pm\frac{4a_0 }{b^2v}+\mathcal{O}(b_0^3,a_0b_0,a_0^2)~,\label{Teo-4} \\
&& \delta\hat\alpha_{\mathrm{T},r}\approx\frac{1}{4}\left(\frac{1}{r_R^2}+\frac{1}{r_S^2}\right) \left[bb_0+b_0^2\pm a_0/v\right]~.\label{Teo-finite}
\end{eqnarray}
It is seen that to the leading nontrivial order of $1/r_R$ and $1/r_S$, the effect of finite distance becomes universal to other parameters in the deflection of the rays. We plot this correction in Fig. \ref{Figure4} (b) for $a_0=b_0^2, ~b=10^2b_0^2,~r_R=10^4b_0$ and $v=0.9c$ as a function of $r_S$. Again, as expected, this finite-distance correction also monotonically decreases as $r_S$ increases, and similar to the Kerr case in Eq. \eqref{Kerr-finite}, the direction of the wormhole rotation has little effect on this correction in this range of $r_S$. 

\section{Conclusion and discussions} \label{CONCLU}
In the weak field approximation, we have studied the gravitational deflection of massive particles for a receiver and source at finite distance from stationary, axisymmetric and asymptotically flat lens. For this purpose, we have extended the generalized optical metric method to generalized Jacobi metric method according to the JMRF metric. By the definition of deflection angle in Refs.~\cite{OIA2017,OIA2018,OIA2019} and the GB theorem applying to a quadrilateral with a generalized Jacobi metric, the deflection angle as a global effect is considered and it can be calculated by the sum of two parts: the surface integral of the Gaussian curvature of the generalized Jacobi metric and the path integral of geodesic curvature of the particle ray lied in the generalized Jacobi metric space, as show in Eq.~\eqref{gbdef}.

By generalized Jacobi metric method, we have obtained the deflection angle of massive particles for a receiver and source at finite distance from Kerr BH given in Eq.~\eqref{Kerr-1}, and from Teo wormhole given in Eq.~\eqref{Teo-1}. These results cover the deflection angle of light for a receiver and source at finite distance and the infinite-distance gravitational deflection angle of massive particles in these two spacetimes as special cases.
In the limit $r_R\to b$ or $r_S\to b$, the deflection angle of massive particles in Kerr BH is divergent, but that in the Teo wormhole is finite. In Kerr spacetime, the deflection angle increases as the velocity $v$ decreases, as shown in Fig.~\ref{Figure2}. In Teo wormhole spacetime, the deflection angle of prograde particle increases as velocity increase, whereas that decreases as velocity increases in the retrograde case, as shown in Fig.~\ref{Figure4}. The difference is because the effect of velocity to deflection angle in Teo wormhole spacetime is much smaller than Kerr spacetime. The effects of BH spin, subluminal particle velocity and finite distance to the deflection angle in the Kerr spactime are compared in the microlensing and lensing by supermassive BH cases. It is found that the former has the largest effect, while the relative size of the latter two effects can vary according to the exact value of the particle velocity, source or observer distance and other lensing parameters. 

In Ref.~\cite{Jusufi2019-1}, Jusufi \textit{et. al} have shown that one can distinguish the rotating naked singularities from Kerr-like wormholes by the deflection angles of massive particles. It would be interesting to see if the finite-distance deflection angle can do the same thing. In addition, one can extend the method to investigate the finite-distance deflection of massive charged particles by charged gravitational object such as Kerr-Newman BH. Finally, we plan in the near future to extend these results to more complicated spacetimes such as rotating global monopole.

\acknowledgements
 Z. Li thanks S. Chanda for helpful discussions related to the JMRF metric.

\end{document}